\newcommand{\Vc}{\textsc{VINCIA}\xspace}
\newcommand{\Py}{\textsc{PYTHIA 8}\xspace}

\newcommand{\dd}{\text{d}}
\newcommand{\abs}[1]{\left| #1 \right|}
\newcommand{\order}[1]{\mathcal{O}\left( #1 \right)}

\newcommand{\antc}{a_c}
\newcommand{\alphaS}{\ensuremath{\alpha_S}}

\newcommand{\sAK}{\ensuremath{s_{AK}}\xspace}
\newcommand{\saj}{\ensuremath{s_{aj}}\xspace}
\newcommand{\sij}{\ensuremath{s_{ij}}\xspace}
\newcommand{\sjk}{\ensuremath{s_{jk}}\xspace}
\newcommand{\sik}{\ensuremath{s_{ik}}\xspace}
\newcommand{\sAB}{\ensuremath{s_{AB}}\xspace}
\newcommand{\sjb}{\ensuremath{s_{jb}}\xspace}

\newcommand{\sab}{\ensuremath{s_{ab}}\xspace}

\newcommand{\lt}[1]{\ensuremath{#1}'}

\newcommand{\Qemit}{Q_\text{emit}}
\newcommand{\Qstart}{Q_\text{start}}
\newcommand{\Qhad}{Q_\text{had}}

\newcommand{\trial}{\text{trial}}
\newcommand{\ant}{\text{ant}}
\newcommand{\accept}{\text{accept}}

\newcommand{\eqRef}[1]{eq.~\ref{#1}}

\documentclass{PoS}

\usepackage{amsmath}
\usepackage{xspace}

\title{The Vincia Parton Shower}

\ShortTitle{The Vincia Parton Shower}

\author{Walter~T.~Giele$^a$, Lisa~Hartgring$^b$, David~A.~Kosower$^c$, Eric~Laenen$^{b, d, e}$, Andrew~J.~Larkoski$^f$,
Juan~J.~Lopez-Villarejo$^{c, g, h}$, \speaker{Mathias~Ritzmann}$^c$, Peter~Skands$^g$\\
\llap{$^a$}Theoretical Physics Department
Fermilab, P.O. Box 500, Batavia, IL 60510, USA\\
\llap{$^b$}Nikhef, Theory Group, Science Park 105, 1098 XG, Amsterdam, The Netherlands\\
\llap{$^c$}Institut de Physique Th\'eorique, CEA Saclay, F--91191 Gif--sur--Yvette cedex, France\\
\llap{$^d$}ITFA, University of Amsterdam, Science Park 904, 1018 XE, Amsterdam, The Netherlands\\
\llap{$^e$}ITF, Utrecht University, Leuvenlaan 4, 3584 CE, Utrecht, The Netherlands\\
\llap{$^f$}Center for Theoretical Physics, Massachusetts Institute of Technology, Cambridge, MA 02139, USA\\
\llap{$^g$}Theoretical Physics, CERN CH-1211, Geneva 23, Switzerland\\
\llap{$^h$}Univ. Autonoma de Madrid and IFT-UAM/CSIC, Madrid 28049, Spain
}

\abstract{
We summarize recent developments in the \Vc parton shower.
After a brief review of the basics of the formalism, the extension of \Vc to hadron collisions is sketched.
We then turn to improvements of the efficiency of tree-level matching by making the shower history unique and by incorporating identified helicities.
We conclude with an overview of matching to one-loop matrix elements.
}

\FullConference{XXI International Workshop on Deep-Inelastic Scattering and Related Subjects\\
		 22-26 April, 2013\\
		 Marseilles, France}

\begin{document}

\section{Introduction}

The \Vc parton shower \cite{Giele:2007di, Giele:2011cb, GehrmannDeRidder:2011dm, LopezVillarejo:2011ap, Ritzmann:2012ca, Larkoski:2013yi, Hartgring:2013jma} is a plugin to the \Py \cite{Sjostrand:2007gs} event generator based on the dipole-antenna picture of QCD radiation \cite{Gustafson:1987rq}.
It emphasizes comprehensive uncertainty estimates and efficient matching to fixed-order matrix elements (obtained from \cite{Alwall:2007st} in the present implementation).

\section{Formalism}

We give a very condensed overview of the \Vc formalism as presented in \cite{Giele:2007di, Giele:2011cb}.
Like every parton shower, \Vc implements the unitary evolution of a parton configuration from a scale $\Qstart$ on the order of the 
hard scale of the partonic interaction to a scale $\Qhad$ on the order of the hadron masses.
Working in the large $N_C$ limit of QCD, a configuration is organized into pairs of color-connected partons, called  
antenn\ae{}, which emit independently.
The algorithm can then be characterized by the non-emission probability when evolving an antenna from $\Qstart$ to $\Qemit$, 
also referred to as the Sudakov factor:
\begin{equation}
	\Delta \left( Q_\text{start}^2, Q_\text{emit}^2 \right) 
= \exp \left[ - \int_{\Qemit^2}^{\Qstart^2} a_c \, \dd \Phi_\ant \right], \quad
a_c = g_s^2 \left( Q^2 \right) \, C \, \bar{a} , 
\label{AntennaNonEmission}
\end{equation}
where $a_c$ is the antenna function composed of the coupling constant, the color factor and the color- and coupling-stripped antenna function $\bar{a}$. It encodes the singular unresolved limits of tree-level matrix elements\footnote{Neglecting spin correlations and color-subleading contributions.} such that
\begin{equation}
	\label{eq:antc}
	\frac{ \abs{ M_n^0 }^2 }{ \sum_{ijk \to IK} a_c \left(i, j, k \right) \abs{ M_{n-1}^0  \left( \dotsc, I, K, \dotsc \right) }^2 }
	\xrightarrow{\text{single unresolved limit}} 1 \, .
\end{equation}
The antenna is integrated over the antenna phase space $\dd \Phi_\ant$ which satisfies
\begin{equation}
	\dd \Phi_n \left( \dotsc, i, j, k, \dotsc \right) = \dd \Phi_{n-1} \left( \dotsc, I, K, \dotsc \right) \dd \Phi_\ant \left( I, K \to i, j, k \right)
\end{equation}
where $p_i + p_j + p_k = p_I + p_K$ and all momenta are on-shell. In terms of invariants it is given by
\begin{equation}
	\dd \Phi_\ant = \frac{1}{16 \pi^2} \frac{1}{m_{IK}^2} \dd \sij \dd \sjk \frac{ \dd \phi }{2 \pi}
\end{equation}
where $\sij = 2 \, p_i \cdot p_j$ and $\phi$ parametrizes rotations around $\mathbf{p}_I$ in the center-of-mass frame of $p_I+p_K$.
Several choices for the definition of $Q$ are implemented in \Vc, among them $Q^2 \propto \sij \sjk / m_{IK}^2$ and $Q^2 \propto \min(\sij,\sjk)$.

The direct inversion of the Sudakov factor for the emission scale $\Qemit$ is cumbersome.
Therefore, \Vc uses the veto algorithm.
It uses a trial Sudakov factor
\begin{equation}
	\label{eq:SudaTrial}
	\Delta^\trial \left( \Qstart^2, \Qemit^2 \right) = \exp \left[ - \int_{\Qemit^2}^{\Qstart^2} a_c^\trial \dd \Phi_\ant^\trial \right]
\end{equation} 
to generate trial emissions, where $a_c^\trial$ is an overestimate of the physical antenna function and 
$\Phi_\ant^\trial$ is a hull of the physical antenna phase space, both chosen such that the inversion of \eqRef{eq:SudaTrial} is simple.
The trial emissions are then accepted with probability 
\begin{equation}
	\label{eq:Paccept}
	P_\accept = \frac{ a_c }{ a_c^\trial } \left[ \{ p \} \in \Phi_\ant \right]
\end{equation}
where the angle bracket denotes unity if the condition is true and zero otherwise.

\section{Extension to Hadron Collisions}

To extend the \Vc{} formalism to incoming hadrons, as described in more detail in \cite{Ritzmann:2012ca}, we have to include the ratio of parton distribution functions before and after the branching in the integrated antenna \cite{Sjostrand:1985xi}:
\begin{equation}
  \Delta ( Q_\text{start}^2, Q_\text{emit}^2 ) = 
  \exp \left[ - 
  \int_{Q_\text{emit}^2}^{Q_\text{start}^2}
  \antc\, \frac{ f_a (x_a, Q^2 ) }{ f_A (x_A, Q^2 ) } 
     \frac{ f_b (x_b, Q^2 ) }{ f_B(x_B, Q^2) }
	\dd \Phi_\text{ant} \right] \,,
\label{AntennaIntegralISR}
\end{equation}
where $a$, $b$ are the generic labels we use for incoming partons after branching and $A$, $B$ denote their pre-branching counterparts.
The resolution is defined in \cite{Ritzmann:2012ca} as the crossing of $Q^2 \propto \abs{\sij} \abs{\sjk}/ (\abs{\sij} + \abs{\sjk} + \abs{\sik})$.
This amounts to an evolution variable which does not change if an outgoing parton is replaced by an incoming one with 
the same momentum. Alternative definitions of $Q$ for incoming partons are left to future investigations.

The antenna functions $a_c$ still obey \eqRef{eq:antc}, but since incoming partons cannot become soft, they are not the crossings of their final-final counterparts in general.

We also need to specify the phase space factorization used if incoming partons are involved.
At variance with the final-final case, only the integral of the phase space over the momentum fraction(s) of the incoming emitter(s) factorizes.
For initial--final antenn\ae{} we use a variant which does only 
rescale the incoming emitter~\cite{Catani:1996vz}:
\begin{equation}
  \label{eq:factorizationif}
 	\int \frac{ \dd x_a}{ x_a} \dd \Phi_3 \left( -a,-b ; j, k, R \right) = 
	\\
	\int \frac{ \dd x_A }{ x_A } \dd \Phi_2 \left( -A, -b ; K, R \right) 
	\\
	\dd \Phi_\text{ant}^{if} \left(-A;K \to -a; j, k \right)
\end{equation}
where $R$ denotes one or more outgoing particles and incoming partons are indicated by a minus sign.
 The pre- and post-branching momentum fractions 
are related by $x_A/x_a = \sAK/(\sAK+\sjk)$.
The initial--final antenna phase space in terms of invariants is
\begin{equation}
\dd \Phi_\text{ant}^{if} \left(-A;K \to -a; j, k \right) = 
	\\
	\frac{1}{ 16 \pi^2} \frac{ \sAK }{ (\sAK+\sjk)^2 } \dd \sjk \dd \saj 
\end{equation}
with the boundaries $0 \leq \sjk \leq \sAK (1-x_A)/x_A$, 
$0 \leq \saj \leq \sAK + \sjk$.
The third coordinate of the initial--final phase space has been suppressed above since the antenna function does not depend on it.

Initial--initial branchings add a parton with momentum transverse to the beam to the event.
In the phase space factorization we use, this transverse momentum is compensated by 
all the other particles in the event (both colored and colorless).
This is implemented by replacing $p_A, p_B$ by $p_a, p_b, p_j$, with $p_A + p_B = p_a + p_b - p_j$, followed by 
applying a Lorentz boost such that $p_a$ and $p_b$ are aligned with the beams again.
The corresponding phase space factorization is~\cite{Catani:1996jh}
\begin{equation}
\int \frac{ \dd x_a }{x_a} \frac{ \dd x_b }{x_b} 
	\dd \Phi_2 \left( -a, -b ; j, R \right) 
	\\
	= \int \frac{ \dd x_A }{x_A} \frac{ \dd x_B }{x_B} 
	\dd \Phi_1 \left( -\lt{A}, -\lt{B} ; \lt{R} \right)
	\dd \Phi_\text{ant}^{ii}
\end{equation}
with the initial--initial antenna phase space in terms of invariants given by
\begin{equation}
\dd \Phi_\text{ant}^{ii}\left(- \lt{A}, -\lt{B}   \to -a, -b ; j \right) = \\
		 \frac{1}{16 \pi^2} \frac{\sAB}{ \sab^2} 
		  \theta \left( 1- x_a \right) \theta\left( 1-x_b \right) \dd \saj \dd \sjb
\end{equation}
where we have suppressed the integration over the angle $\phi$ 
parametrizing rotations around the beam.
Momenta $X'$ are Lorentz transformed with respect to the frame in which $p_a$ and $p_b$ are aligned with the beam directions.

\section{Uncertainties}

\Vc offers variations of many of its parameters to enable an assessment of the uncertainty its predictions have.
For example, it allows changing the antenna functions $\bar{a}$ by terms which do not contribute in single unresolved limits 
or switching between color factors which differ by terms suppressed by $N_C^{-2}$.
As elaborated upon in \cite{Giele:2011cb}, it also provides the option to explore these uncertainty variations at a small computational cost.
For this purpose, a weight corresponding to each uncertainty variation is produced along with each (unweighted) event. Applying these weights (which average to unity by construction), one obtains the distribution of events as it would have been produced when running \Vc with the modified parameters.
This procedure generalizes naturally to parton distribution function uncertainties for the ratio occurring in \eqRef{AntennaIntegralISR}.

\section{Tree-level Matching}

\Vc implements a unitary matching to consecutive tree-level matrix elements over all of phase 
space\footnote{Technically, there is still a matching scale, starting from the third emission by default. 
However, this scale it is not a defining feature of the matching. It is needed because the shower misses some subleading logarithms.}. 
This means that it produces unweighted events such that the expansion of the probability density for $n$-jet events 
to Born level agrees with the fixed-order description.
This is achieved by changing the accept probability in \eqRef{eq:Paccept} to
\begin{equation}
	P_\accept = \frac{a_c}{a_\trial} \left[ \{ p \} \in \Phi_\ant \right] P_\text{ME}, \quad P_\text{ME}=
	\frac{ \abs{M_n^0}^2 }{ \sum_{ijk \to IK} a_c \left(i, j, k \right) \abs{ M_{n-1}^0  \left( \dotsc, I, K, \dotsc \right) }^2} \, .
\end{equation}
The sum in the denominator of the matching factor $P_\text{ME}$ runs over all $n-1$-parton configurations from which 
the shower could have arrived at the $n$-parton configuration in question.
The formula is valid under the assumption that tree-level matching has already been performed at multiplicity $n-1$ and 
that the shower algorithm does not depend on how we have arrived at a configuration.
By virtue of \eqRef{eq:antc}, the matching factor tends to unity in single unresolved limits.

Using a shower algorithm in which all antenn\ae{} contribute over all of phase space, the cost of calculating 
the denominator of $P_\text{ME}$ grows linearly with $n$.
In \cite{LopezVillarejo:2011ap}, a variant of the \Vc algorithm was presented in which there is always exactly one 
shower history for each phase space point, enabling substantially faster running times for the generation of tree-level 
matched samples.
A further extension of the \Vc formalism to identified parton helicities has been described in \cite{Larkoski:2013yi}, 
building on earlier work in \cite{Larkoski:2009ah}.
It enables to perform the matching by evaluating only individual helicity amplitudes, which is particularly helpful at high multiplicities.

\section{One-loop Matching}

In \cite{Hartgring:2013jma}, matching of \Vc to next-to-leading order has been described 
and demonstrated explicitly for to $Z \to 3 \text{ jets}$.
The paradigm is the same as for tree-level matching: A differential matching factor is applied to 
the trial accept probability such that the expansion of the parton shower agrees with the fixed-order result.

In the the example of $Z \to \text{ 3 jets}$, we want to reproduce the exclusive 3-jet rate:
\begin{equation}
	R_3^\text{ex} = 
	\abs{ M_3^0 }^2 + 2 \Re \left( M_3^0 M_3^{1*} \right) + \int_0^{Q_\text{had}^2} \frac{ \dd \Phi_4 }{ \dd \Phi_3 } \abs{ M_4^0 }^2
\end{equation}
where it is required that additional radiation is unresolved at the scale $Q_\text{had}$ at which the shower is stopped.
For this purpose, a fully differential matching factor is required such that
\begin{equation}
	(1 + V_3) \text{Approximate} = R_3^\text{ex} + \order{ \alphaS^3 }
\end{equation}
where ``Approximate'' stands for the tree-level matched shower approximation.
$V_3$ is written almost completely in terms of integrals which are calculated analytically.
The only integrals entering $V_3$ which have to be computed numerically are related to the fact that due to the 
tree-level matching at the four-parton level, $\abs{M_4^0}^2$ enters into the calculation of $V_3$.
However, these integrals are well-behaved and give small contributions for almost all phase space points.

\section{Acknowledgements}

The work presented here was supported in part
by the U.S. Department of Energy under contract No.~DE-AC02--07CH11359,
by the Netherlands Foundation for Fundamental Research of Matter (FOM) programme
104, entitled ``Theoretical Particle Physics in the Era of the LHC'', 
by the National Organization for Scientific Research (NWO),
by the European Research Council under Advanced Investigator Grant ERC--AdG--228301,
by the US Department of Energy under cooperative research agreement DE--FG02--05ER41360,
by the U.S. National Science Foundation, grant NSF--PHY--0969510, 
by the European Commission (HPRN--CT-- 200-00148), FPA2009--09017 (DGI del MCyT, Spain) and S2009ESP--1473 (CA Madrid),
and by a MEC grant, AP2007--00385.

\end{document}